\documentclass[graybox, secnum]{svmult}

\usepackage{mathptmx}       
\usepackage{helvet}         
\usepackage{courier}        
\usepackage{type1cm}        
\usepackage{makeidx}         
\usepackage{graphicx}        
\usepackage{multicol}        
\usepackage[bottom]{footmisc}
\usepackage{hyperref}        
\usepackage{soul}            
\hypersetup{colorlinks=true,urlcolor=blue}
\usepackage[square,numbers]{natbib}
\newcommand{\hbindex}[1]{{\sethlcolor{green}\hl{#1}}\index{#1}}  
\makeindex             

\usepackage[exponent-product=\cdot,separate-uncertainty=true]{siunitx}  
\usepackage{enumitem}

\definecolor{darkblue}{rgb}{0,0,0.8}
\definecolor{darkred}{rgb}{0.6,0,0}
\hypersetup{
    linkcolor=darkred,          
    citecolor=darkblue,        
}
\renewcommand{\hbindex}[1]{#1}

\begin{document}

\title*{DEPFET Active Pixel Sensors}
\author{Norbert Meidinger\thanks{corresponding author} and Johannes M\"uller-Seidlitz}
\institute{Dr. Norbert Meidinger \at Max Planck Institute for Extraterrestrial Physics, Gie\ss{}enbachstra\ss{}e 1, 85748 Garching, Germany, \email{norbert.meidinger@mpe.mpg.de}
\and Dr. Johannes M\"uller-Seidlitz \at Max Planck Institute for Extraterrestrial Physics, Gie\ss{}enbachstra\ss{}e 1, 85748 Garching, Germany, \email{johannes.mueller-seidlitz@mpe.mpg.de}}
%
%
\maketitle

\abstract{An array of DEPFET pixels is one of several concepts to implement an active pixel sensor. Similar to PNCCD and SDD detectors, the typically \SI{450}{\um} thick silicon sensor is fully depleted by the principle of sideward depletion. They have furthermore in common to be back-illuminated detectors, which allows for ultra-thin and homogeneous photon entrance windows. This enables relatively high quantum efficiencies at low energies and  close to \SI{100}{\%} for photon energies between \SI{1}{\keV} and \SI{10}{\keV}. Steering of the DEPFET sensor is enabled by a so-called Switcher ASIC and readout is performed by e.g. a VERITAS ASIC. The configuration enables a readout time of a few microseconds per row. This results in full frame readout times of a few milliseconds for a $512 \times 512$ pixel array in a rolling shutter mode. The read noise is then typically three electrons equivalent noise charge RMS. DEPFET detectors can be applied in particular for spectroscopy in the energy band from \SI{0.2}{\keV} to \SI{20}{\keV}. For example, an energy resolution of about \SI{130}{\eV}~FWHM is achieved at an energy of \SI{6}{\keV} which is close to the theoretical limit given by Fano noise. Pixel sizes of a few tens of microns up to a centimetre are feasible by the DEPFET concept.}


~\\

\section{Introduction}
\label{sec:intro}

The DEPFET concept had already been proposed by Kemmer and Lutz in 1987 \cite{kemmer87}. Since then, various implementations were developed for different applications and scientific fields. They serve as particle trackers in the high radiation environments near the collision point of particle accelerators. Mounted at an X-ray free-electron laser, they can help to unveil the nature of fast processes on tiny scales in various disciplines.

The first spectroscopic DEPFETs for space applications were developed for the Mercury Imaging X-ray Spectrometer aboard the BepiColombo mission orbiting the planet Mercury \cite{treis10}. The next space project with the employment of DEPFETs is the Wide Field Imager (WFI) of ATHENA, ESA's Advanced Telescope for High-Energy Astrophysics \cite{nandra13}.

These silicon-based DEPFET concepts have been designed and the devices fabricated in the Semiconductor Laboratory of the Max-Planck-Society \cite{lutz05}. They are similar to those of PN-implanted Charge Coupled Devices (PNCCD) \cite{meidinger21} and Silicon Drift Detectors (SDD). All three detector types feature a sideward depletion \cite{gatti84} which enables a sensitivity over the full chip thickness. Therefore, all three sensor types can be back illuminated which allows for a thin and homogeneous photon entrance window over the sensor area. In PNCCDs, the signal charge needs to be transferred along the channel to a readout node, which is not necessary for DEPFETs. An SDD detector has a very high time resolution in the order of a microsecond but comprises typically a small number of large cells. The generated signal is readout immediately and not stored. In a DEPFET, each pixel has a transistor implemented for charge storage and signal amplification as well as a second transistor for the signal charge clear afterwards.

\section{Detector Concept}
\label{sec:concept}

An active pixel sensor like the DEPFET features the first signal amplification already inside each pixel of the imaging sensor. Since such a readout node is implemented in every pixel, there is no need of charge transfer and no risk of potential charge loss due to traps in a transfer channel. As a drawback, the complexity of the detector is increased significantly because every pixel needs two transistors and the necessary steering and readout contacts.

The implementation and the readout concept will be explained first. The photon detection and charge collection as well as the electronics necessary to steer and read out the sensor are described afterwards.

The starting point for the sensor is a thin slice (wafer) of monocrystalline silicon. Such a semiconductor can be doped to influence its resistivity characteristics. Dopants are atoms with a number of outer shell electrons---the valence electrons---that differs from the four electrons in a silicon atom’s outer electron shell. If silicon atoms in the crystal lattice are replaced by such a dopant atom, the additional electron or the missing one, a hole, change the electrical properties. A region with arsenic or phosphorus dopants---elements with five valence electrons---is called n-doped. The additional, weakly bound electrons with their negative charge are the majority charge carriers while the holes are called minority charge carriers. Doping with boron---an element with three valence electrons---leads to weakly bound holes, positively charged quasiparticles, that are the majority charge carriers in such a p-doped region. In both cases, the opposite charge in the atomic nucleus is stationary. The overall electric charge of a doped region is neutral.

At a p-n junction, where a p-doped and an n-doped region adjoin each other in the same crystal, the majority charge carriers of both regions diffuse into the other region and recombine. The stationary dopant atoms remain with four electrons. Thus, they are electrically charged. An electrical field between the now positive dopants in the n-doped region and the negative dopant atoms in the p-doped region is established. The resulting drift and the diffusion act in opposite directions and thereby an equilibrium is set up. The volume without majority charge carriers is called space charge region. By applying an external voltage, the space charge region can be extended (reverse bias, no current is possible) or it can be shrunken, even down to zero (forward bias, current in one direction).

For a more precise and deeper introduction into the terms introduced above, especially about the band model in solids, it is referred to the literature \cite{kittel04}.

\subsection{DEPFET Principle}
\label{sec:depfet}

An X-ray photon interacts in silicon and generates electrons, which drift to the storage region underneath the transistor in the centre of a pixel. The resulting transistor current increase is measured and, after calibration, gives the X-ray photon energy.

As transistor, a \hbindex{DEPFET} is used. It is a DEpleted P-channel Field Effect Transistor, which performs amplification and switching. For a \hbindex{MOSFET}---a Metal Oxide Semiconductor FET---a source and a drain region are implanted into the silicon wafer material. For the DEPFET, these are strong p-implants (p+) in initially slightly n-doped (n-) material. Space charge regions are formed around the source and drain regions. The surface is covered with an isolating layer, typically silicon dioxide and silicon nitride. Between the source and the drain implants on top of the isolating layer, a metallic contact is placed---the transistor gate \cite{kahng60}. In case of the DEPFET, it is formed of polycrystalline silicon. With a sufficiently negative voltage at the gate, holes will be collected below. They form a conductive layer, the p-channel, between source and drain. The assignment of the p+ implant being source or drain is performed via proper bias voltages. Since there is a hole current in the p-channel, the source is the contact with the more positive voltage level. In addition to such a simple MOSFET, a shallow n-doping is implanted below the transistor channel for the DEPFET. It is the potential minimum for electrons collected in the sensitive volume of the sensor. To avoid recombination and to facilitate the collection of signal electrons, the entire sensing device is depleted by applying a sufficient high reverse bias to remove all majority charge carriers. The collected electrons generate mirror charges that are additional holes in the conductive channel which increase the conductivity of the transistor channel proportional to the number of collected electrons. The region is called Internal Gate because its function is similar to that of the (external) transistor gate. This increase of the transistor current allows for the determination of the X-ray photon energy.

The so-called charge gain $g_q$ quantifies the change in the current between source and drain $I_{DS}$ as a function of the collected signal charge $Q_{sig}$. For constant voltages between source and drain ($U_{DS}$) as well as source and gate ($U_{GS}$) the charge gain is proportional to square root of the current $I_{DS}$ between source and drain.
\begin{equation}\label{eq:gq}
    g_q = \frac{\delta I_{DS}}{\delta Q_{sig}} \propto \sqrt{\frac{2 \mu_h}{W L^3 C_{ox}} I_{DS}}
\end{equation}
$C_{ox}$ is the capacity per unit area of the gate oxide, $L$ the length and $W$ the width of the gate, $\mu_h$ the hole mobility \cite{lutz99}. Due to the fact that the number of mirror charges is smaller than the number of collected electrons, the charge gain is proportional and not equal to the right term of \autoref{eq:gq}.

\begin{figure}[t]
\sidecaption
\includegraphics[width=7.49cm]{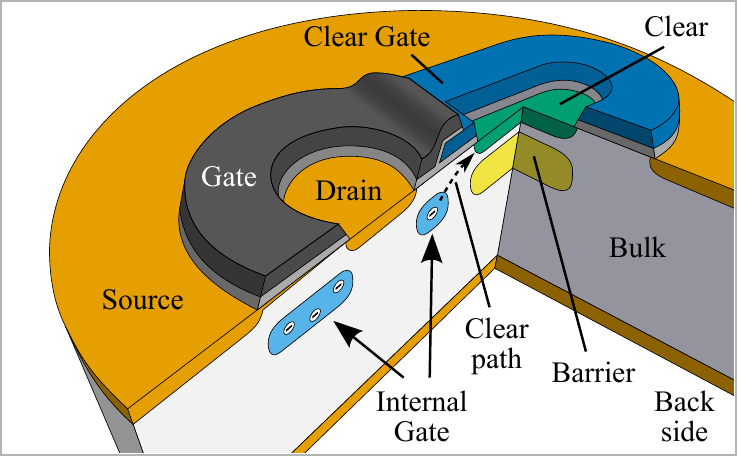}
\caption{The DEPFET concept. Electrons collected in the Internal Gate increase the conductivity of the transistor channel between source and drain proportionally to their number. Afterwards, the collected electrons are removed by the clear transistor. The barrier below the clear shields the electrons in the bulk from a drift directly into the clear contact.}
\label{fig:depfet}       
\end{figure}

To remove the collected charge carriers from the Internal Gate, a second transistor of NMOS type is implemented which allows for a controlled drift of the electrons towards the positive clear contact. A barrier below the clear contact shields it against the bulk to avoid the drift of signal electrons after their generation directly to the clear contact (see \autoref{fig:depfet}). A DEPFET pixel comprises both transistors, an adapted PMOS FET for signal sensing and the NMOS FET for the charge reset afterwards. To build an imaging detector, an appropriate pixel array is created. If pixels larger than \SI{50}{\um} are needed, drift rings can be added around the transistors similar to a silicon drift detector (SDD)~\cite{lutz99}. By a full depletion of the sensor thickness, back-illumination is feasible (see \autoref{sec:charge}). For a typical thickness of \SI{450}{\um}, a reverse bias in the order of \SI{100}{V} is necessary at the back side. 

\subsection{Photon Interaction}
\label{sec:photon}

\begin{figure}[t]
\sidecaption
\includegraphics[width=7.49cm]{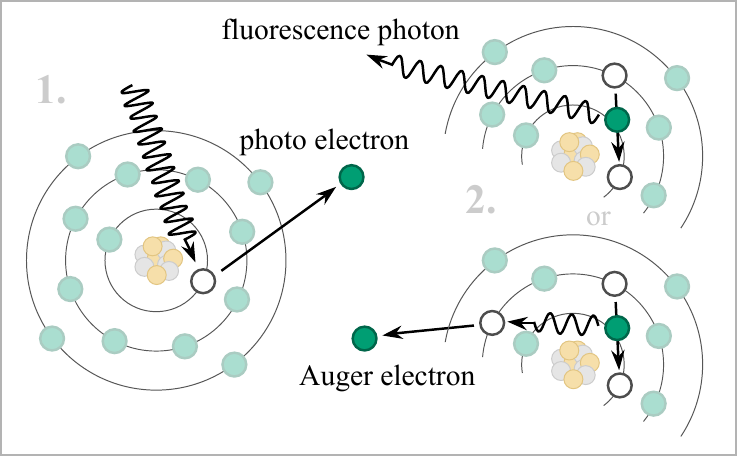}
\caption{First most likely interactions of an absorbed X-ray photon with an energy below a few ten kiloelectron volts. The photon ionises a silicon atom. The empty position is then occupied by an electron from an outer shell. The surplus energy is emitted via a photon. This photon can ionise the same atom again and a so-called Auger electron is ejected from the atom.}
\label{fig:interact}       
\end{figure}

When an X-ray photon of up to an energy of a few ten kiloelectron volts interacts with the silicon sensor, typically a photo electron is generated. As depicted in \autoref{fig:interact}, a fluorescence photon or an Auger electron are emitted in addition. These processes will continue until thermalisation is reached. This results in a number of signal electrons proportional to the photon energy. The mean energy needed for the generation of an electron-hole pair is about $w = \SI{3.71}{\eV}$ at \SI{200}{\K} \cite{lowe07}. The number of generated electrons varies according to \hbindex{Fano} statistics \cite{fano47} and causes a noise contribution called Fano noise. 
\begin{equation}\label{eq:fano}
    \sigma_{\text{Fano}} = \sqrt{F w E} = \sqrt{0.118 \cdot \SI{3.71}{\eV} \cdot E}
\end{equation}
with $\sigma_{\text{Fano}}$ as the standard deviation, $F = 0.118$ the material specific Fano factor for silicon \cite{lowe07} and E the energy of the incident photon in electron volts.

\subsection{Charge Collection}
\label{sec:charge}

\begin{figure}[t]
\sidecaption
\includegraphics[width=7.49cm]{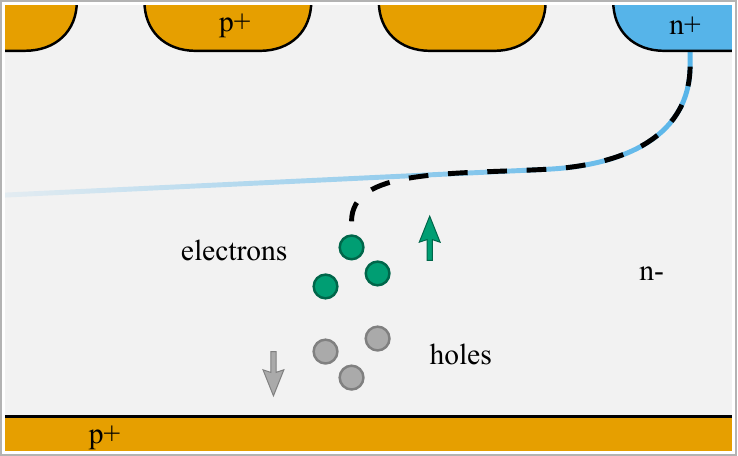}
\caption{Charge collection in a sidewards depleted sensor \cite{gatti84}. Electrons drift along the electric field towards the potential minimum (blue). In vertical direction, an asymmetric, parabolic, electric potential separates the electrons and holes. A sideward drift is induced by a static gradient of the applied voltages at the front side p+ implants at the top.}
\label{fig:sidedepl}       
\end{figure}

The concept of full depletion enables X-ray photon detection over the entire chip thickness of \SI{450}{\um} \cite{gatti84}. This allows for a high quantum efficiency up to more than \SI{15}{\keV}. For this purpose, a negative voltage of about \SI{-100}{\V} has to be applied at the photon entrance window in order to create a drift field for electrons towards the front side of the sensor. To enable an immediate sideward drift in large pixels, a structured front side in combination with a gradient in the bias voltages is necessary as shown in \autoref{fig:sidedepl}. The electrons generated by an incident X-ray photon drift along the electric field towards the potential minimum in each pixel. There, in the Internal Gate, the signal electrons are stored until they are measured and afterwards cleared.

A further advantage of full depletion is the possibility to have the photon entrance window at the back side of the sensor chip. It can be realised as an ultrathin layer with an uniform thickness over the entire sensor area. The structures at the front side would prevent such a layer (see \autoref{fig:depfet}). A thin layer is required to reduce the region where generated electrons recombine with holes and are lost for the signal measurement. Due to their lower absorption length, this is in particular important for low energetic X-ray photons.

\subsection{Steering and Readout Electronics}
\label{sec:asics}

\begin{figure}[t]
\sidecaption
\includegraphics[width=7.49cm]{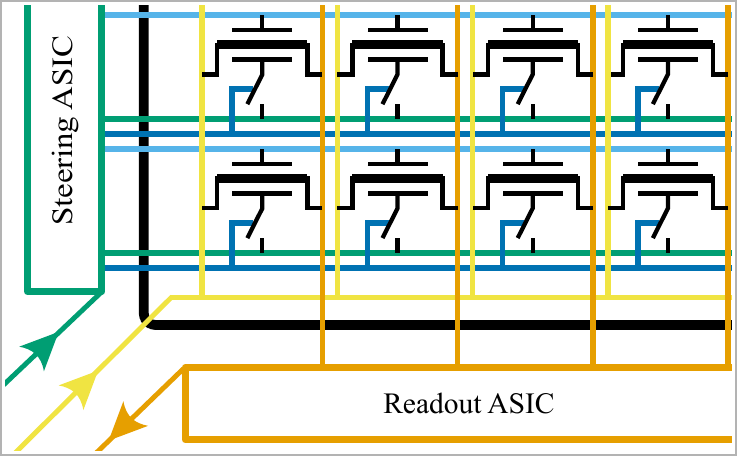}
\caption{Equivalent circuit diagram for a DEPFET detector. Four by two pixels are shown. The gates (connected via light blue lines), clears (green) and clear gates (dark blue) are connected to the steering ASIC. The sources are biased globally (yellow) and the drains are read out via the readout ASIC (orange).}
\label{fig:ecd}       
\end{figure}

As shown in \autoref{fig:ecd}, the pixels of the current DEPFET X-ray sensors are connected row-wise to steering ASICs (Application Specific Integrated Circuit), the so-called Switcher \cite{fischer03}. The switchable contacts of a DEPFET (gate, clear gate and clear) are supplied with appropriate voltage levels for on- and off-state of the DEPFET transistors. For the readout of the signal charges collected in the DEPFET pixels, a second type of ASIC is needed. It performs the further amplification and shaping. By use of a multi-channel readout ASIC like the \hbindex{VERITAS} \cite{porro14}, a DEPFET column is connected to a dedicated ASIC channel. Thereby, the signals of these channels are processed simultaneously. After the processing is completed, the individual voltage signals are serialised to an output buffer. By this method, just one ADC (Analogue to Digital Converter) is needed per readout ASIC.

\section{Operation}
\label{sec:operation}

To reduce the complexity of the detector system and its power consumption, a DEPFET detector for space applications is typically operated in a rolling shutter mode. The Switcher ASIC turns on just one row of the DEPFET sensor while all other rows are switched off. Switched off pixels are still collecting incoming signal charges but consume no power. After the readout by the VERITAS is finished, the row is switched off and the next one is switched on and read out. This will be continued until all the rows of a frame are read out and the sequence starts again with the next frame.

\begin{figure}[t]
\sidecaption
\includegraphics[width=7.49cm]{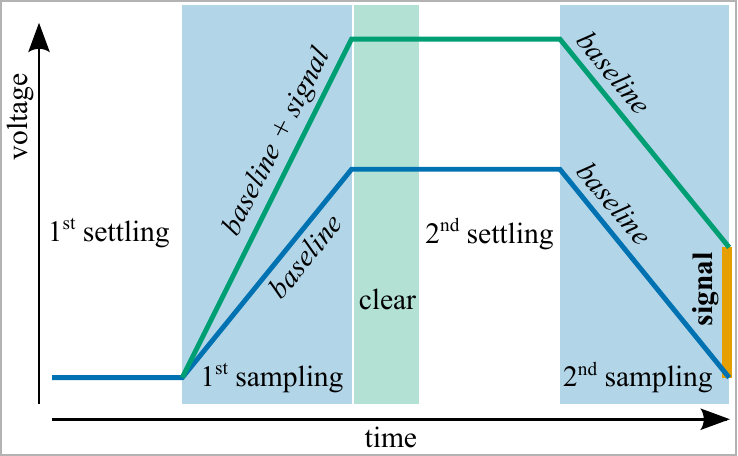}
\caption{Readout scheme of a spectroscopic X-ray DEPFET with (green) and without (blue) signal. It consists of a first sampling, a clear of the signal electrons and a second, inverted sampling. The difference between the two samplings gives the signal and is proportional to the collected charge. Settling times need to be introduced to stabilise the signal level.}
\label{fig:readout}       
\end{figure}

The readout itself is split into two parts: In a first step, the baseline signal of the pixel is read together with the collected electrons in the Internal Gate. Then, the electrons are removed from the Internal Gate via the clear transistor. Afterwards, a second signal sampling of the empty pixel is done. The VERITAS ASIC calculates the difference between the two levels (see \autoref{fig:readout}). This correlated double sampling is applied to reduce the read noise.

Each electron collected in the Internal Gate of a DEPFET pixel increases the voltage level in the ASIC by a certain amount. The voltage difference measured by the readout ASIC gives thus the photon energy after calibration.

The main operating voltages for the operation of a DEPFET sensor are typically as follows:
\begin{itemize}[leftmargin=4em]
\item{Source: \SI{0}{\V} (reference voltage)}
\item{Drain: \SI{-6}{\V} to \SI{-3}{\V}}
\item{Gate: \SI{-2}{\V} (on) and $>\SI[retain-explicit-plus]{+3}{\V}$ (off)}
\item{Clear Gate: \SI[retain-explicit-plus]{+5}{\V} to \SI[retain-explicit-plus]{+10}{\V} (on) and \SI{0}{\V} (off)}
\item{Clear: \SI[retain-explicit-plus]{+15}{\V} to \SI[retain-explicit-plus]{+20}{\V} (on) and \SI[retain-explicit-plus]{+1}{\V} (off)}
\item{Back Contact: \SI{-100}{\V}}
\end{itemize}
For larger pixels with drift structures, additional voltages at the front side may be needed.

To achieve a higher time resolution, the readout can be optimised. The VERITAS ASIC allows for a parallel signal processing and serialisation of the previously processed row. In addition, only a part of the sensor, a window, can be read out while the rest of the frame is discarded. The readout time scales with the number of rows inside the window \cite{ms18b}.

Optimal spectroscopy requires a sufficiently low dark current. Thus, cooling of the DEPFET sensor is necessary, typically below \SI{-40}{\celsius}. Furthermore, a constant temperature is needed for precise calibration (\autoref{sec:calibration}).

\section{Performance Characteristics}
\label{sec:performance}

The spectral performance is dominated by the Fano noise for photon energies above \SI{1}{\keV}. At lower photon energies, the recombination of electron-hole pairs at the entrance window and thus signal loss contributes significantly to the energy resolution. The use of DEPFETs is best suited for an energy range of \SI{0.2}{\keV} to \SI{15}{\keV}. At a line energy of \SI{0.2}{\keV}, the spectrum is still of Gaussian shape. The quantum efficiency (QE) at \SI{10}{\keV} is \SI{96}{\%} and at \SI{15}{\keV} it is still \SI{63}{\%}. The QE at low energies depends largely on the need for an optical blocking filter which can be deposited directly on the entrance window. A further key performance parameter is a read noise of about 3 electrons ENC (Equivalent Noise Charge) RMS. The pixel size which determines the spatial resolution can be matched in the range from about \SI{50}{\um} to centimetre scale to the angular resolution of the optics. The time resolution scales with the number of sensor rows. A typical readout time per row is in the order of a few microseconds.

\subsection{Energy Resolution}
\label{sec:enres}

The energy resolution of a DEPFET detector is not only determined by the Fano noise and the read noise which includes the dark current contribution. Additional noise contributions are given by:
\begin{itemize}[leftmargin=2em]
    \item Charge losses due to recombination of electron-hole pairs close to the entrance window where the separating electric field is weak.
    \item Incomplete charge collection because of a drift of signal electrons directly to the clear contact instead of to the Internal Gate.
    \item Energy misfits describe signal read losses \cite{ms18b}. These can be caused by charge clouds arriving in the Internal Gate during the first signal sampling (see \autoref{fig:readout}) or at the end of the clear pulse.
    \item The signal charge can be split over more than one pixel. For energy determination, the signal values have to be recombined. In space missions, the telemetry rate is limited and thus, data reduction is necessary. For this purpose, an event threshold needs to be set to discriminate between signal and noise events. Small signal fractions below the event threshold are lost for recombined events.
    \item Instrument background mimicking X-ray source photons is caused by interaction of cosmic rays with the detector surrounding material. It consists typically of secondary electrons and photons \cite{eraerds21}. The background can be reduced by the Self-Anti-Coincidence technique \cite{grant20} that detects heavily ionising particles with the DEPFET detector itself and requires no additional Anti-Coincidence detector.
\end{itemize}
All these effects result in a broadening of a spectral line and apart from the background also in a signal loss. The energy resolution is parameterised by the Full Width at Half Maximum (FWHM) of the line. The shape of the spectrum is described by the detector response function which is energy dependent and accounts for the various charge loss effects.

\begin{figure}[t]
\sidecaption
\includegraphics[width=7.49cm]{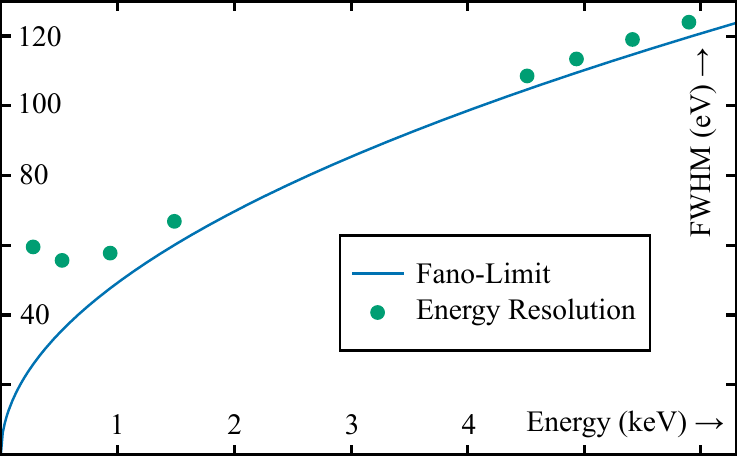}
\caption{Energy resolutions for different emission lines. From lower to higher energies, these are: C~K-$\alpha$, O~K-$\alpha$, Cu~L-$\alpha$, Al~K-$\alpha$, Ti~K-$\alpha$, Ti~K-$\beta$, Cr~K-$\alpha$, Mn~K-$\alpha$. Entrance window effects degrade the spectral performance at lower energies. The measurements were taken with a $256 \times 256$ pixel prototype sensor for ATHENA's WFI at a readout speed of \SI{6.1}{\us} per row \cite{treberspurg19}.}
\label{fig:fwhm}       
\end{figure}

In \autoref{fig:fwhm}, the energy resolution of different emission lines is shown. For higher energies, the measured FWHMs are close to the Fano limit while at energies below \SI{1}{keV} the above mentioned effects degrade the spectral performance \cite{treberspurg19}.

\subsection{Performance Degradation in Space}

There are three potential effects that can reduce the detector performance in space.

Radiation damage is caused by protons and alpha particles. They can destroy the silicon lattice which results in an increase of the dark current. To mitigate radiation damage effects, an appropriate shielding thickness depending on the orbit and mission lifetime and sufficient low DEPFET temperatures are necessary to lower the thermal generation current. An advantage of an active pixel sensor like the DEPFET compared to a Charge Coupled Device (CCD) is that no charge transfer is needed. Radiation damage effects in CCDs typically affect the charge transfer efficiency and thus the energy resolution even if the average transfer loss is corrected. Soft protons focused by the mirror on the focal plane do not cause critical radiation damage on the back-illuminated DEPFET. The reason is, that the transistors are accommodated on the front side.

A second potential performance degradation is caused by contamination of the photon entrance window. Molecules from inside or outside of the instrument can accumulate on the cold sensor surface. As a result the QE is reduced, primarily for low photon energies. The standard mitigation strategy is a bake-out of all instrument and satellite components before assembly to minimise outgassing in space. A further prevention measure is a warm optical blocking filter in front of the detector as the higher temperature minimises the accumulation of contamination from outside. A cold trap with a temperature below the one of the sensor could also reduce molecular contamination on the sensor and filter.

\begin{figure}[t]
\sidecaption
\includegraphics[width=11.7cm]{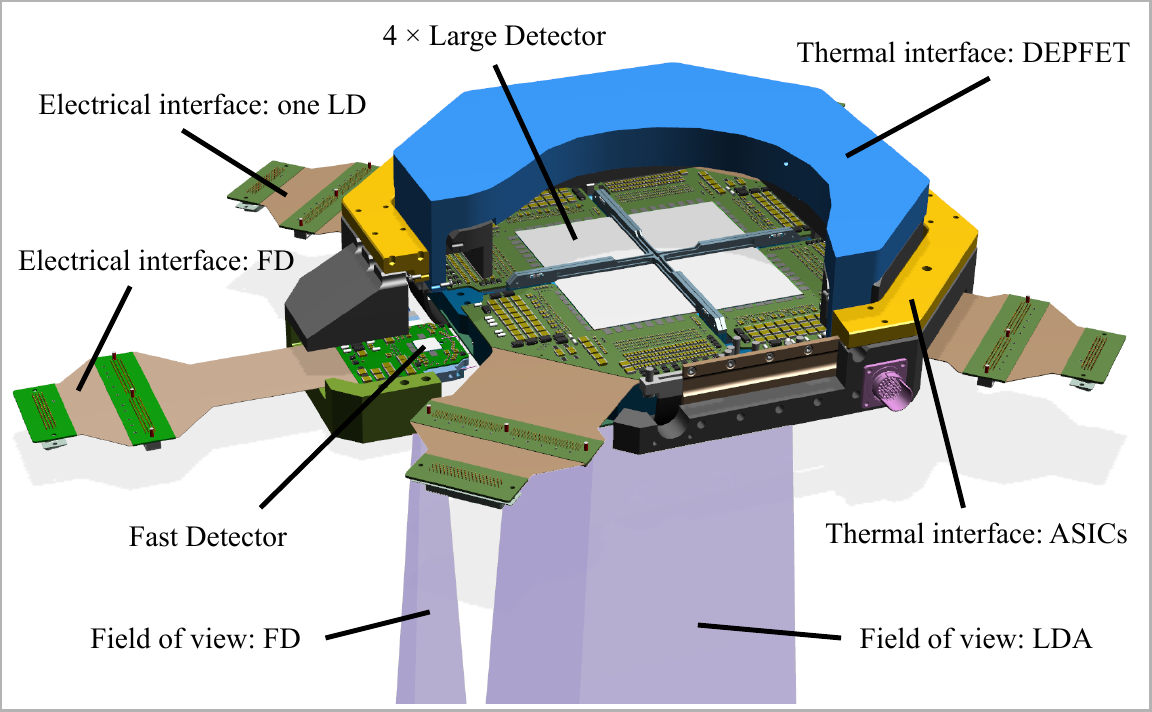}
\caption{The drawing shows the WFI detector assembly including the mechanical structure and the thermal architecture. The Fast Detector (FD) has an image area of $\SI{8.3}{\mm} \times \SI{8.3}{\mm}$ whereas the Large Detector Array (LDA) has a sensitive area of $\SI{133}{\mm} \times \SI{133}{\mm}$.}
\label{fig:ch}       
\end{figure}

With the increasing mirror area of nowadays missions, the probability that micro-meteoroids are deflected on the sensors rises as well. When a micro-meteoroid hits a pixel, the dark current can increase by a large amount and the spectroscopic performance of the hit pixel is deteriorated. An advantage of a DEPFET and its fast readout is the short time of dark current accumulation. A further mitigation is the choice of a low operating temperature.

\section{Example Case: ATHENA WFI detector}
\label{sec:athena}

The novel DEPFET detector concept was for the first time used in space for the Mercury Imaging X-ray Spectrometer (\hbindex{MIXS}) on \hbindex{BepiColombo}, a satellite for the exploration of planet Mercury \cite{bunce20} realised as joint mission of ESA and JAXA. The purpose of the MIXS instrument is the analysis of the elemental composition of the planet's surface by imaging X-ray spectroscopy of the fluorescence lines \cite{majewski14}. The DEPFET sensors consist of $64 \times 64$ pixels with a pixel size of $\SI{300}{\um} \times \SI{300}{\um}$. They are steered by Switcher ASICs and read out by two Asteroid readout ASICs \cite{porro10}. This enables a readout time per frame of \SI{170}{\us}. The required energy resolution is $\le \SI{200}{\eV}$~FWHM at \SI{1}{\keV} energy. BepiColombo was launched in October 2018 and will reach an orbit around Mercury end of 2025. The MIXS detector performance was verified during the travel to Mercury. The measured energy resolution was \SI{139}{\eV}~FWHM at an energy of \SI{5.9}{\keV} \cite{bunce20}.

\begin{figure}[t]
\sidecaption
\includegraphics[width=7.49cm]{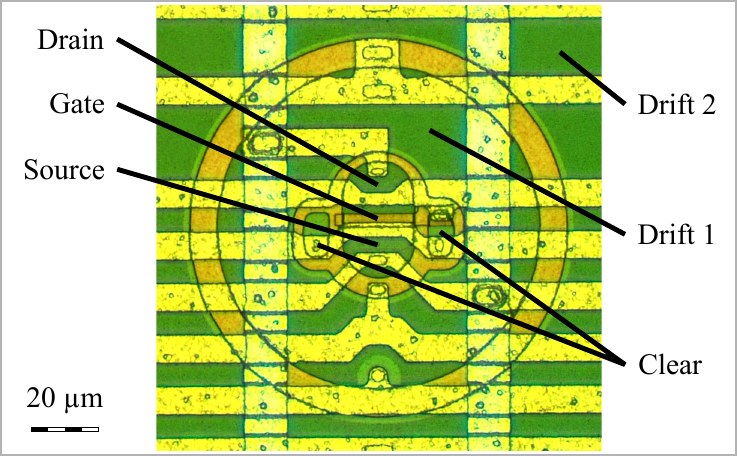}
\caption{Photo of an ATHENA WFI  pixel ($\SI{130}{\um} \times \SI{130}{\um}$). In the centre, the DEPFET with the two clear contacts is accommodated. Two drift structures focus the charge carriers to the Internal Gate located below the (external) transistor gate. A metal supply line grid connects the pixel contacts to the ASICs and the global supplies (see \autoref{fig:ecd}).}
\label{fig:pixel}       
\end{figure}

The next application for DEPFETs in space is planned for the \hbindex{ATHENA} \hbindex{Wide Field Imager} (WFI), one of two focal plane instruments. ATHENA is ESA's next generation large class X-ray mission \cite{nandra13}. The launch to the first Lagrange point of the Sun-Earth system is scheduled for 2034. The WFI instrument is designed for imaging and spectroscopy over a large field of view of $\SI{40}{\arcmin} \times \SI{40}{\arcmin}$ and high count rate observations of up to and beyond 1 Crab source intensity \cite{meidinger14, meidinger20}. The WFI focal plane comprises therefore two complementary and independent detectors: A large detector array (LDA) consisting of four quadrants (LD) with $512 \times 512$ pixels each and a fast detector (FD) with $64 \times 64$ pixels, operated in split frame readout to improve the time resolution (see \autoref{fig:ch}). The pixel size of both detectors is $\SI{130}{\um} \times \SI{130}{\um}$, which matches the envisaged angular resolution of the silicon pore X-ray optics of \SI{5}{\arcsecond} half energy width (HEW). A photo of such a pixel is shown in \autoref{fig:pixel}.

At the beginning of the project, a technology development was performed to identify the optimum transistor design and technology for the WFI instrument \cite{treberspurg18a}. To fulfil the timing requirements, a linear transistor gate design was selected (see \autoref{sec:options}). Furthermore, the chosen thin gate oxides allow for self-aligned implants that enable a better uniformity of the pixel performance.

The ATHENA science cases require a time resolution of $\le \SI{5}{\ms}$ for the large detector array and $\le \SI{80}{\us}$ for the fast detector. In addition, an energy resolution of \SI{170}{\eV}~FWHM at an energy of \SI{7}{\keV} and \SI{80}{\eV}~FWHM at \SI{1}{\keV} energy is required until the end of life. The overall detection efficiency needs to achieve \SI{4.4}{\%} at \SI{0.2}{\keV}, \SI{80}{\%} at \SI{1}{\keV}, \SI{93}{\%} at \SI{7}{\keV} and \SI{90}{\%} at \SI{10}{\keV} for both detectors.

\begin{figure}[t]
\sidecaption
\includegraphics[width=7.49cm]{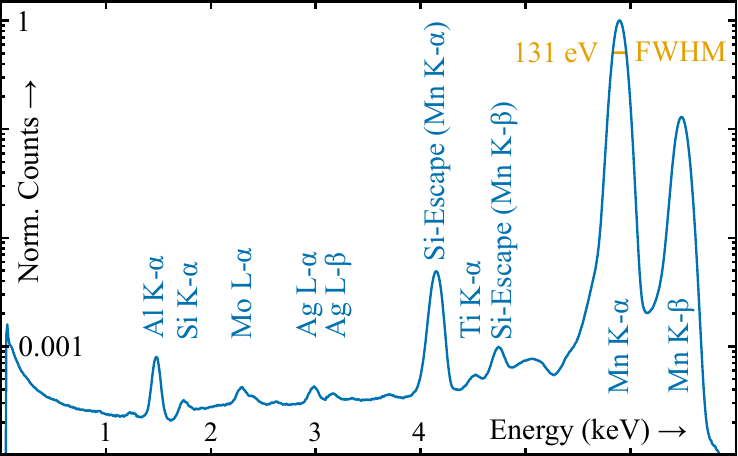}
\caption{\textsuperscript{55}Fe spectrum measured with a $512 \times 512$ pixel DEPFET developed for ATHENA's WFI. The FWHM of the Mn K-$\alpha$ line at \SI{5.9}{\keV} is \SI{131}{\eV} for a readout time of the entire sensor of \SI{2}{\ms} and a DEPFET temperature of \SI{-60}{\celsius}. In addition, the Mn K-$\beta$ line at \SI{6.5}{\keV}, the associated Si-escape lines and various fluorescence lines appear.}
\label{fig:spectrum}       
\end{figure}

In \autoref{fig:spectrum}, an energy spectrum measured with a large DEPFET detector of $512 \times 512$ pixels is shown. Apart from the dominant emission lines generated by the \textsuperscript{55}Fe source, further peaks appear. First, the silicon escape peaks which are caused by an escape of a silicon K-$\alpha$ fluorescence photon (see \autoref{fig:interact}) from the sensitive sensor volume. Thereby, the signal of the Mn~K photon is reduced by the $\SI{1.7}{\keV}$ of the silicon escape photon. Second, fluorescence lines are observed which are generated when source photons hit material in the vicinity of the detector.

\begin{figure}[t]
\sidecaption
\includegraphics[width=11.7cm]{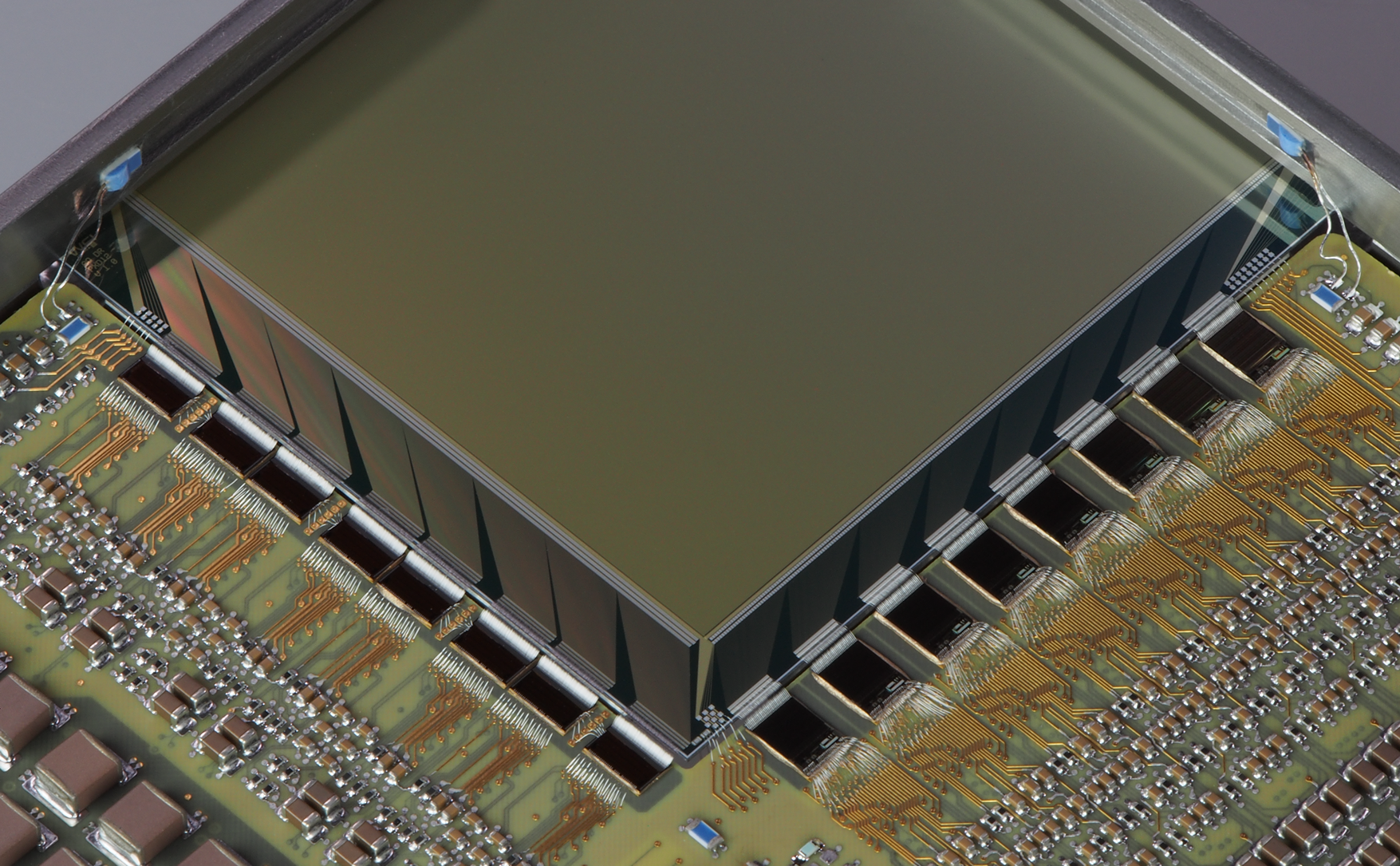}
\caption{Large DEPFET detector with $512 \times 512$ pixels. Eight Switcher steering ASICs (left) with 64 channels each select the rows to be switched on and trigger the reset in the DEPFET. The eight 64-channel VERITAS ASICs (right) read the signal from the pixels that are switched on by the Switcher ASICs. The photon entrance window is located on the opposite side.}
\label{fig:ld}       
\end{figure}

\begin{figure}[t]
\sidecaption
\includegraphics[width=11.7cm]{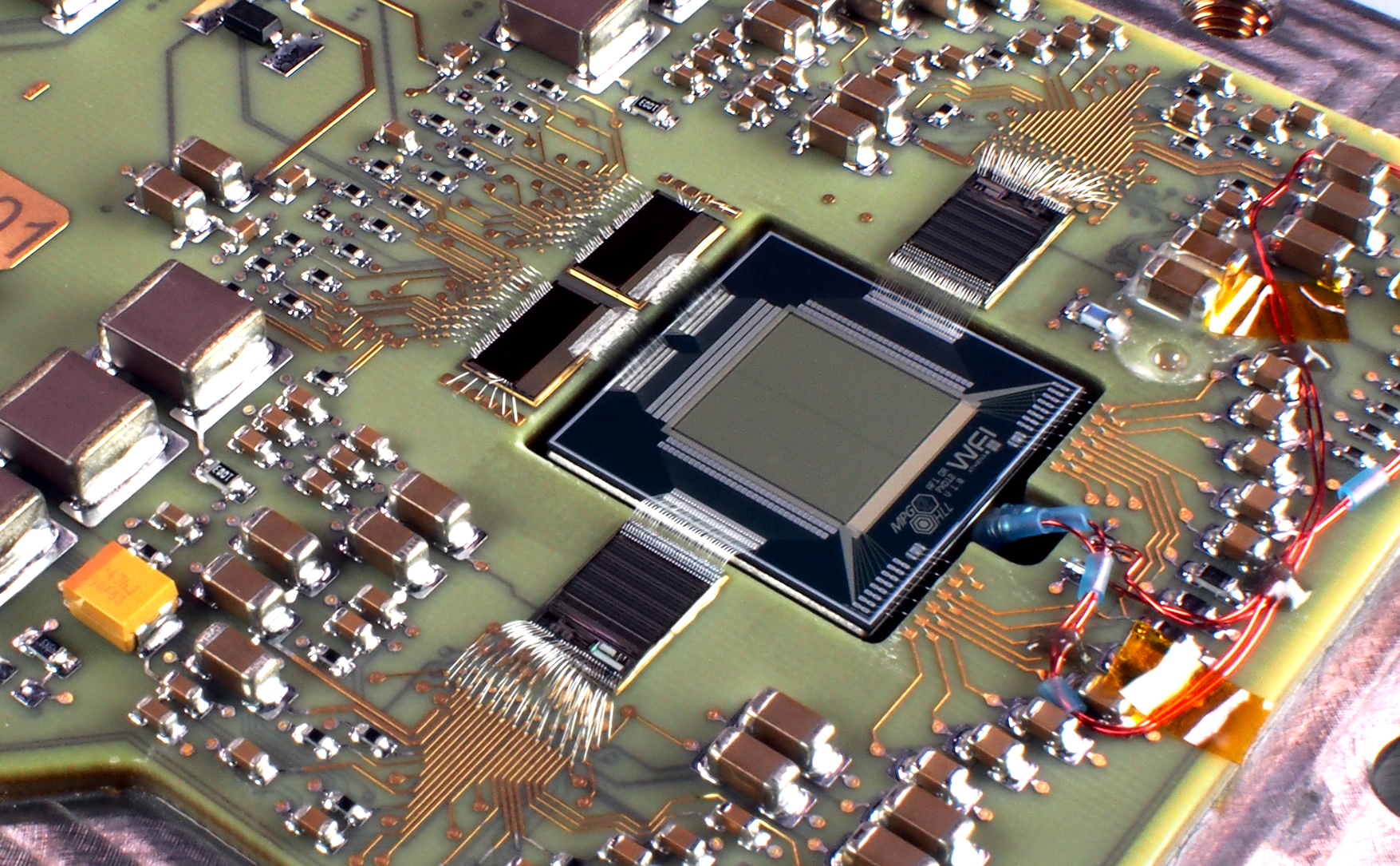}
\caption{The Fast Detector's front side. The readout of the sensor is split into two parts and, therefore, is surrounded by the two steering and the two readout ASICs as well as the corresponding printed circuit board with its electronic components.}
\label{fig:fd}       
\end{figure}

\autoref{fig:ld} shows a quadrant of the LDA and \autoref{fig:fd} the FD. To fulfil the performance requirements, the sensor needs to be cooled down to a temperature range between \SI{-80}{\celsius} and \SI{-60}{\celsius}. The lowest temperature is necessary to meet the energy resolution requirements until the end of the mission. The reason is that the DEPFET thermal generation current increases over mission time due to radiation damage. The front end electronics including the ASICs are operated at a higher temperature to minimise the radiator area on the satellite. The WFI camera uses passive cooling via radiators only. While the power consumption of the FD is \SI{3}{\W}, the LDA dissipates \SI{44}{\W} due to the large number of 64 ASICs that are needed for the fast readout of more than one million pixels. The detectors are connected via flexible leads to the electronics boxes. There, the necessary supply voltages and the timing sequence are generated. In total, 43 supply voltages and 22 steering signals per LD and nearly the same amount for the FD are required. In addition, the analogue output from the VERITAS ASICs are digitised and processed in the detector electronics box. This includes basic pixel signal corrections, e.g. subtraction of a dark image, and event detection to reduce the amount of data generated by $>\SI{260}{\mega pixels\per\second}$ for the detectors that needs to be transmitted to ground. Including all further required signals like housekeeping or the programming interface of the VERITAS ASIC, connectors with more than 200 pins are needed for each LD and FD to be operated by the electronics boxes.

\begin{figure}[t]
\sidecaption
\includegraphics[width=11.7cm]{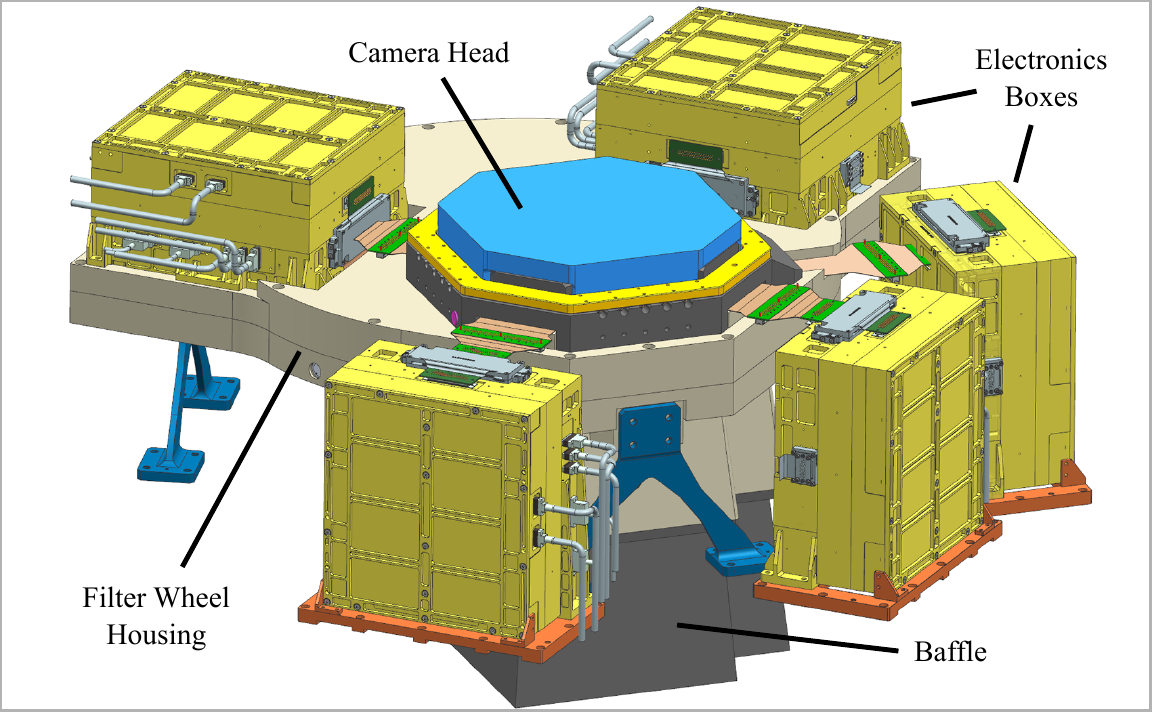}
\caption{The WFI instrument with the camera head, that houses the detectors of FD and LDA. A filter wheel including the optical stray-light baffle is mounted in front of it. Each detector has its dedicated electronics box for supply, signal digitisation and event pre-processing. The bipods (dark blue) are the mechanical interfaces to the instrument platform of the satellite.}
\label{fig:wfi}       
\end{figure}

In front of the detector, a filter wheel is accommodated (\autoref{fig:wfi}). It provides the following functionalities. First, optical blocking filters made of aluminium deposited on polyimide \cite{barbera18}. Second, an onboard calibration source based on a radioactive \textsuperscript{55}Fe source. Third, an open position which allows for observations with higher QE at low energies of optical faint objects. Fourth, a closed position to measure the instrumental background without source photons.

\section{Calibration}
\label{sec:calibration}

Due to the fact, that each pixel has its own readout node, a pixel-wise calibration of the gain is necessary. This requires sufficient high photon statistics for every pixel in each operating mode (full frame, window modes).

The QE depends strongly on the photon energy, in particular for low energies below \SI{0.5}{\keV} and for absorption edges of optical blocking filters and the photon entrance window materials. The photon entrance window of the DEPFET sensor consists of aluminium (optional, as optical blocking filter), silicon nitride and silicon oxide. The total thickness is in the order of a tenth of a micrometre.

The non-linearity of signal and photon energy needs to be calibrated in particular for low energies. The reasons are event detection threshold effects for split events and the electron-hole pair recombination near the photon entrance window.

For the determination of the detector response, measurements of a series of monochromatic lines from the lowest to the highest energy are necessary. Instead of a line spectrum, a continuous spectrum is obtained due to the effects mentioned in \autoref{sec:performance} and \autoref{fig:spectrum}. Based on these measurements, the detector energy response function is developed.

On ground, calibration measurements are typically performed at a synchrotron. In space, regular recalibrations may be necessary due to changes caused e.g. by radiation damage and contamination. The in-orbit calibration can be performed by either an onboard calibration source (e.g. \textsuperscript{55}Fe and stimulated fluorescence lines) or well known, bright cosmic sources.

Calibration in space is an ongoing activity to ensure an optimum detector performance during mission lifetime. Contamination would affect the QE and radiation damage degrades the energy resolution globally and accumulates continuously. In contrast, micro-meteoroid impacts occur as single events that cause a localised, single pixel damage in case of a DEPFET.

\section{Outlook for DEPFET Options}
\label{sec:options}

The implementation of a DEPFET as introduced in \autoref{sec:depfet} can be approached in many ways. The usage of a circular gate as shown in \autoref{fig:depfet} reduces the necessary structures to a minimum. The separation of the source and the drain regions is realised only by the gate itself. One clear contact on one side serves as reset node. This DEPFET design is used for the MIXS instrument onboard BepiColombo.

\subsection{Linear Gate Layout}

\begin{figure}[t]
\sidecaption
\includegraphics[width=7.49cm]{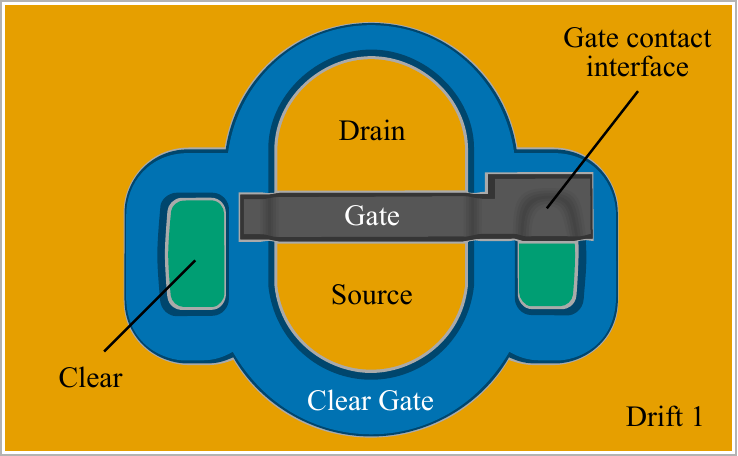}
\caption{DEPFET with a linear gate as it is implemented for ATHENA's WFI (see also \autoref{fig:pixel}). The linear gate design allows for narrower gates. Due to the on average shorter distance to the clear, the reset of the internal gate charge is improved. The gate contact interface for the contact hole placed on the side enables shorter gate lengths in addition.}
\label{fig:linear}       
\end{figure}

However, the circular gate layout sets limits to its gate dimensions and, thereby, to the achievable spectral performance in combination with high readout speeds \cite{treberspurg18b}. The size of the drain limits the gate width to at least \SI{40}{\um} with the current technology. This sets also constraints on the distance an electron has to drift during the clear process. A long drift distance requires a long clear time. In addition, the contact for the gate needs to be placed directly onto the gate structure, which limits the gate length to a minimum of \SI{5}{\um}. By shifting to a linear gate design, the gate width and length can be further reduced. The design enables the introduction of a contact interface above a clear transistor. Therefore, the gate length is independent of the minimum contact hole size to the metal supply grid. The reduced transistor channel area results in an increased amplification of collected electrons (see \autoref{eq:gq}) and a lower input capacitance. Both parameters predominately determine the noise of the detector system \cite{treberspurg18b}. As a drawback, two clear contacts have to be used, one at each end of the gate as shown in \autoref{fig:linear}. Since the clear contacts are regions of potential charge loss, it may degrade the performance. However, for a back illuminated device the effect should be negligible because signal charge generation next to the clear contact is very unlikely. For such a layout, source and drain have no immanent separation. Therefore, an extended clear gate structure was used for the pixel layout of ATHENA's WFI. This avoids the implementation of additional contacts and ensures a proper charge clearing.

\subsection{Prevention of Energy Misfits}

\begin{figure}[t]
\sidecaption
\includegraphics[width=7.49cm]{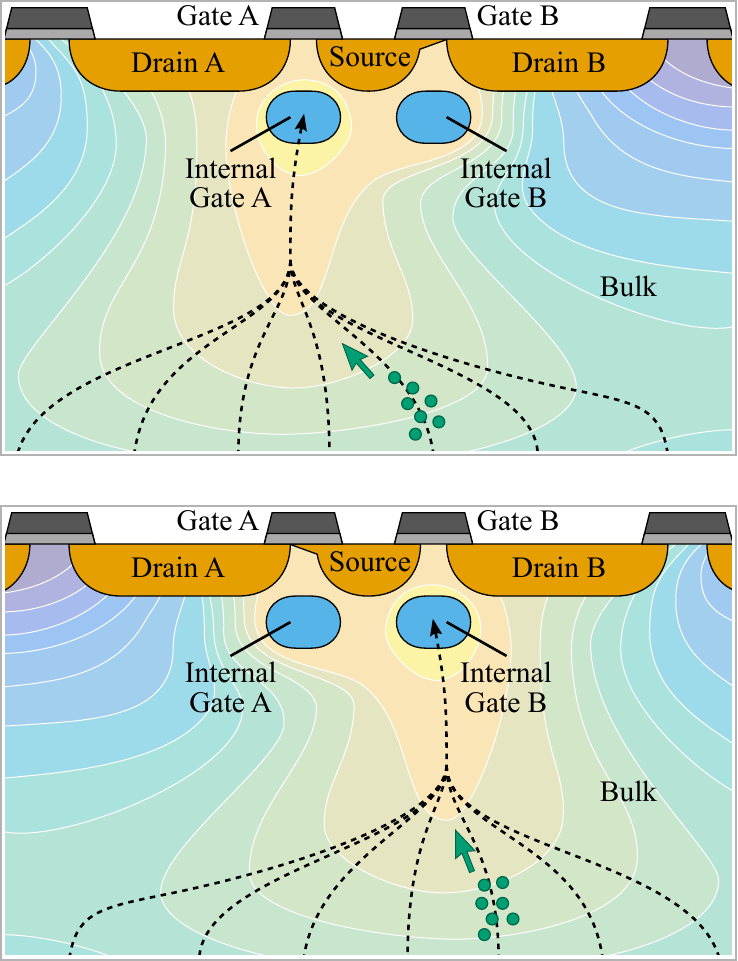}
\caption{DEPFET with two Internal Gates to avoid energy misfits. In the upper figure, sub-pixel~B is read out via the common source contact in a frame n. Sub-pixel~A is used to collect the incoming electrons in the meanwhile. The sub-pixels' functionalities (readout and collection) are changed after every frame defined by the two switchable drain voltages. Therefore, in the frame n+1 (lower figure), sub-pixel~B collects electrons while sub-pixel~A is read out. By switching the global drain voltages, the beginning of a frame is defined for all pixels at the same time. Charge clouds split between two frames can be recombined in a subsequent analysis. In principle, this avoids any charge loss due to the readout process.}
\label{fig:infinipix}       
\end{figure}

The best time resolution in the order of a few microseconds can be achieved using a full parallel readout of all pixels simultaneously \cite{ms18b}. This requires a direct coupling of each pixel to a readout channel. With an increasing fraction of the readout time to the total exposure time, the fraction of energy misfits (see \autoref{sec:enres}) can rise up to almost \SI{50}{\%} and degrade the spectral performance significantly. The use of a shutter would result in a substantial loss of photons and is thus not the optimum solution. To avoid the occurrence of a large fraction of misfits, the charge collection and the readout regions need to be decoupled by spatial separation. One implementation of such a concept is the so-called Infinipix \cite{baehr18}. Each pixel is subdivided into two sub-pixels as shown in \autoref{fig:infinipix} with a common source contact. The two Internal Gates act as potential minima for electrons either for the charge collection or for the readout. The sub-pixels' functionalities are defined via the drain voltages. The drain of the collecting sub-pixel is as positive as the source contact and thus much more attractive to signal electrons. The drain voltages are switched for all pixels simultaneously and define the beginning of a new frame. Therefore, every sub-pixel is read out only every second frame.

The working principle has been successfully demonstrated \cite{baehr14, ms18a}. Already for the operation in rolling shutter mode, small Infinipix test matrices of $32 \times 32$ pixels showed a noticeable increase in the performance. The FWHM at \SI{5.9}{\keV} energy could be improved from \SI{131.4}{\eV} to \SI{125.6}{\eV} by using the Infinipix concept instead of a standard DEPFET. In a three row window mode, the difference is even more significant. Compared to a standard DEPFET, the FWHM at \SI{5.9}{\keV} energy was enhanced from \SI{144.4}{\keV} to \SI{131.2}{\keV}. The spectral performance of the Infinipix is degraded for the three row window mode measurement because in the preliminary data analysis events split between two frames were not recombined \cite{ms18b}.  In near future, a matrix wired to be read out fully parallel will be tested.

\section{Conclusion}
\label{sec:conclusion}

The DEPFET active pixel sensor offers a novel detector concept for future X-ray missions. Meanwhile, the design and technology are so mature to allow for application in a space mission. A first DEPFET detector was launched into space onboard the BepiColombo satellite in 2018. The next application will be the Wide Field Imager of ESA's ATHENA X-ray observatory. The challenging requirements of the ATHENA mission, e.g. high time resolution, led to a further improvement of the DEPFET concept. As a result, an optimal DEPFET transistor design has been developed for ATHENA and large-scale sensors have been manufactured and successfully tested. The DEPFET concept provides a high flexibility to optimise the detector parameters for individual mission objectives like even higher time resolution or scalable pixel sizes from a few tens of microns up to a centimetre.


\end{document}